\documentclass{article}
\usepackage{arxiv}
\usepackage[utf8]{inputenc} 
\usepackage[T1]{fontenc}    
\usepackage{hyperref}       
\usepackage{url}         
\usepackage{booktabs}    
\usepackage{amsfonts}      
\usepackage{nicefrac}      
\usepackage{microtype}     
\usepackage{lipsum}
\usepackage{float}
\usepackage{graphicx}
\usepackage{caption}
\usepackage{subcaption}
\usepackage{titlesec}
\usepackage{multirow}
\usepackage{amsmath}
\usepackage{soul,color}

\usepackage[nodisplayskipstretch]{setspace}

\title{Active learning based generative design for discovery of wide band gap materials }
\author{
  Rui Xin, Edirisuriya M. D. Siriwardane, Yuqi Song, Yong Zhao, Steph-Yves Louis, Alireza Nasiri \\
  Department of Computer Science and Engineering\\
  University of South Carolina\\
  Columbia, SC 29201 \\
    \And
 Jianjun Hu *\\
 Department of Computer Science and Engineering\\
  University of South Carolina\\
  Columbia, SC 29201 \\
  \texttt{jianjunh@cse.sc.edu} \\
}
\titlespacing {\section}
{0pt}{5.5ex plus 1ex minus .2ex}{3.3ex plus .2ex}
\begin{document}
\maketitle
\begin{abstract}
Active learning has been increasingly applied to screening functional materials from existing materials databases with desired properties. However, the number of known materials deposited in the popular materials databases such as ICSD and Materials Project is extremely limited and consists of just a tiny portion of the vast chemical design space. Herein we present an active generative inverse design method that combines active learning with a deep variational autoencoder neural network and a generative adversarial deep neural network model to discover new materials with a target property in the whole chemical design space. The application of this method has allowed us to discover new thermodynamically stable materials with high band gap (SrYF$_5$) and semiconductors with specified band gap ranges (SrClF$_3$, CaClF$_5$, YCl$_3$, SrC$_2$F$_3$, AlSCl, As$_2$O$_3$), all of which are verified by first principle DFT calculations. Our experiments show that while active learning itself may sample chemically infeasible candidates, these samples help to train effective screening models for filtering out materials with desired properties from the hypothetical materials created by the generative model. The experiments show the effectiveness of our active generative inverse design approach.

\end{abstract}

\section{Introduction}

Computational discovery of new materials with desired functions has big potential in a variety of industries. Currently there are two major categories of approaches including 1) computational screening of existing materials in materials databases \cite{fanourgakis2020universal} to find candidates with new properties and 2) generation of new materials using rational design methods\cite{heo2019composition}, crystal structure prediction \cite{oganov2019structure} or generative machine learning (ML) models \cite{noh2020machine,dan2020generative}. The computational screening approach, however, usually suffers from the limited known materials in terms of both quantity and diversity. This is especially true as most known materials in existing databases such as ICSD, Materials Project\cite{jain2013commentary}, and OQMD\cite{kirklin2015open} have only structural information annotated with a few density functional theories (DFT) \cite{min2020accelerated} calculated primitive properties such as formation energies or band gaps. However, few of them are annotated with more interesting macro properties such as thermal conductivity (<2000) and ion conductivity (<200), which makes it challenging to train a highly generalizable machine learning model for screening. The generation approaches also face a variety of challenges, such as the difficulty to generate diverse novel materials compositions and structures and the same issue of lacking high-quality property data for generated hypothetical materials. Essentially, both ML-assisted computational screening and generation suffer from very limited training data with scarce annotated property characterization data.

To improve the efficiency of materials screening, inverse design methods \cite{zunger2018inverse,tagade2019attribute,chen2020generative,noh2019inverse,kim2020inverse,kim2018deep,chen2020design,jiang2020multiobjective} have been proposed to search materials for a given target functionality, in which a sampling or optimization process is used to select a subset of the whole design space for locating the optimal design candidate. The sampling process in inverse design methods is usually guided by a global optimization algorithm\cite{kim2020inverse,zhang2017computer} together with function performance evaluations via atomic simulation. Global optimization methods used in inverse design can be categorized into several types \cite{elsawy2020numerical} including gradient-based optimization methods, gradient-free optimization algorithms such as genetic algorithms, particle swarm optimization, and Covariance Matrix Adaptation Evolution Strategy (CMA-ES), neural network-based generative approaches, and Bayesian optimization methods using surrogate evaluation models. Depending on the discrete or continuous nature of the design targets, not all these optimization methods can be applied directly. For example, the gradient methods cannot be easily applied to search the inorganic materials composition or crystal structure space as they tend to generate infeasible solutions during optimization processes.

Due to the strong physicochemical and geometric constraints among the atoms in periodic crystal materials, we find that gradient-based search operators or crossover/random mutation-based genetic search operators might still be too random in identifying the next feasible sampling point for expensive atomic simulation to obtain its property value. Instead, this exploration step can be guided by a machine learning model as the previously explored samples can be used to learn the structure to properties relations. The resulting trained surrogate model can then be used to predict the properties of unexplored materials with a speed of 100 orders of magnitude faster than conventional DFT calculation methods. This has led to the emergence of active learning methods for materials design\cite{min2020accelerated,yuan2018accelerated,lookman2019active}.


The basic idea of active learning is its capability to explore the design space actively to search candidate with desired properties or identify informative samples for building property prediction models for labeling so that better prediction models can be trained with the existing and minimal number of newly acquired/labeled samples\cite{settles2009active}. An active learning model is typically composed of the following components: an active learner with a specific machine learning prediction model and a query strategy to recommend new samples for labeling, and an oracle model to label the suggested samples \cite{schroder2020survey}. The oracle model can be experimental or computational validations such as first principle DFT calculation or even machine learning models. In general, active learning is very useful for tasks where data is not sufficient or expensive to acquire, which is the case for materials design problems. Since the active learner actively chooses
the training data points, the number of data points needed for training a model based on active learning is typically lower or much lower than the number used by the passive learner. Active learning algorithms can be categorized by the choices of their components, especially the query strategies, which can be classified into uncertainty-based, ensemble-based, and diversity-based approaches. The key idea of an uncertainty-based method is to query those areas about which predicted value the active sampler is least certain. Uncertainty sampling strategies can be employed not only with probabilistic models, which help sample the instance whose best labeling is the least confident, but also non-probabilistic models, which can be modified to have probabilistic output. Such as nearest-neighbor-based classifiers in which neighbors are voting on class labels, and the ratio of the vote on a targeted class represents the posterior probability\cite{lindenbaum2004selective}. Ensemble-based methods maintain a series of machine learning models and help sampler decide valuable query points with which these models most disagree. ACTIVE-DECORATE \cite{melville2004diverse}, which uses DECORATE committees to select good training examples, outperforms traditional ensemble methods such as Bagging and Boosting. DECORATE Algorithm is designed to use additional artificially generated training data from data distribution and label them different from current ensemble predictions in order to generate highly diverse ensembles. Among all these methods, uncertainty-based active learning has been most successfully applied to materials design.


Despite the success stories of a few active learning based materials design cases, there exist several limitations to improve. For example, in \cite{min2020accelerated}, active learning has been applied to find high band gap materials. Their model is initially trained with only around 2\% of the entire search space, and 20 new samples suggested from the optimization schemes, are added at each optimization cycle. Their experiments show that the active learning approach is able to identify the materials with the highest property value after searching only around 7.0\% and 7.7\% of the entire database. However as a demonstration paper, this work does not identify any new materials and have very limited design space compared to the real-world active inverse design problem. In addition, current active learning for materials design studies have been dominated by screening existing materials databases such as ICSD or Materials Project \cite{min2020accelerated}, which limits their search space to a small portion of the vast chemical design space which may unexpectedly exclude more promising candidates\cite{allahyari2020coevolutionary}. 

We propose a method to combine active learning with generative machine learning to build the active generative inverse design framework for new materials discovery and apply it to the discovery of materials with a desired bandgap. Compared to the existing active learning search of high band gap materials, which is focused on exploring the known database, our study focuses on an active learning-based exploration of the unknown design space for high band gap materials. The framework includes four components, including a materials composition generator, a variational autoencoder sampler, a band gap predictor, a Bayesian optimization-based sampler, and a substitution-based structure predictor. Our composition-based generative machine learning model \cite{dan2020generative} can generate millions of new compositions that satisfy basic chemical rules such as charge neutrality and balanced electronegativity and tend to have low free energy. We then build a composition-based band gap prediction model using a Random Forest regressor. A Gaussian process regressor-based Bayesian optimization algorithm is used to train an ML-based sampling model, which suggests the next sampling points in the active learning/search process to maximize expected property improvement. To verify the predicted materials, we use substitution-based structure predictors to get candidate structures which can then be used to verify their structural stability and their band gaps. Our active generative design (AGD) method has allowed us to find a  high band gap material SrYF$_5$ with a band gap of 6.42 eV and seven semiconductor materials with desired band gap values. 

Our contributions include the following:

\begin{itemize}
    \item We propose a novel active generative inverse design approach for discovering new materials in the whole design space rather than in a given database. We find that samples suggested by active learning model may not be chemically valid. But they are useful for building good property prediction models.
    \item We show that active learning can be used to increase property prediction model's capability to screen out high band gap materials by sampling informative data points for property evaluation, which is highly desirable for screening materials  with small datasets of property annotations. 
    \item Our evaluation experiments show that active learning based sampling can be combined with generative models to speed up new materials discovery by training a ML models from samples traversed by the active learning search process for screening  hypothetical material candidates created by the generative models. 
    \item Our new pipeline has been able to discover eleven semiconductor materials (Table2) and two high band-gap materials (SrYF$_5$, RbSr$_2$ClF$_4$) with a DFT-validated band gap of 6.42 eV and 5.64 eV.
\end{itemize}

\section{Materials and Methods}

\subsection{Active Generative Design (AGD) Framework for Materials Discovery}

The main idea of our AGD framework is to use the active learning approach to learn a highly accurate property prediction model and combine it with a generative model (e.g., an autoencoder model or a Generative Adversarial Network (GAN model)). Our whole active generative design framework is shown in Figure 1. It is composed of three modules: an active learning model for training a high-performance property predictor using limited sample labeling,  and a candidate generation model, and a screening model for the discovery of new materials. For the purpose of fast benchmark studies here, a machine learning-based oracle model is trained to simulate the DFT calculations of band gaps in this study.

The active learning module is the part where data augmentation via active exploration of the design space is performed to acquire labels of the most informative sampling points to improve the performance of the machine learning-based Screening Model. The generative module uses VAE or GAN models to generate candidate materials and feed them to the Screening model trained by the enhanced dataset to filter out a number of promising materials candidate formulas. Their corresponding candidate structures will subsequently be generated by exploiting their similarity with existing structures using the element-substitution approach. Formation Energy will also be predicted using Global Attention Graph Neural Network (GATGNN) \cite{louis2020graph}. 

Within our active generative design framework, there are two ways to discover hypothetical materials:
\begin{itemize}
\item 1) ALS: direct active learning search in the whole chemical space, which is done by combining the active learning sampling model with the autoencoder decoding model.
\item 2) AL+RF/Roost with MATGAN: active learning-based screening of candidates generated by the MATGAN generative model. Active learning is used to create labeled samples for training ML screening models using RF or Roost, which are then used to filter the samples generated by MATGAN.
\end{itemize}


\begin{figure}[ht]
  \centering
  \includegraphics[width=\linewidth]{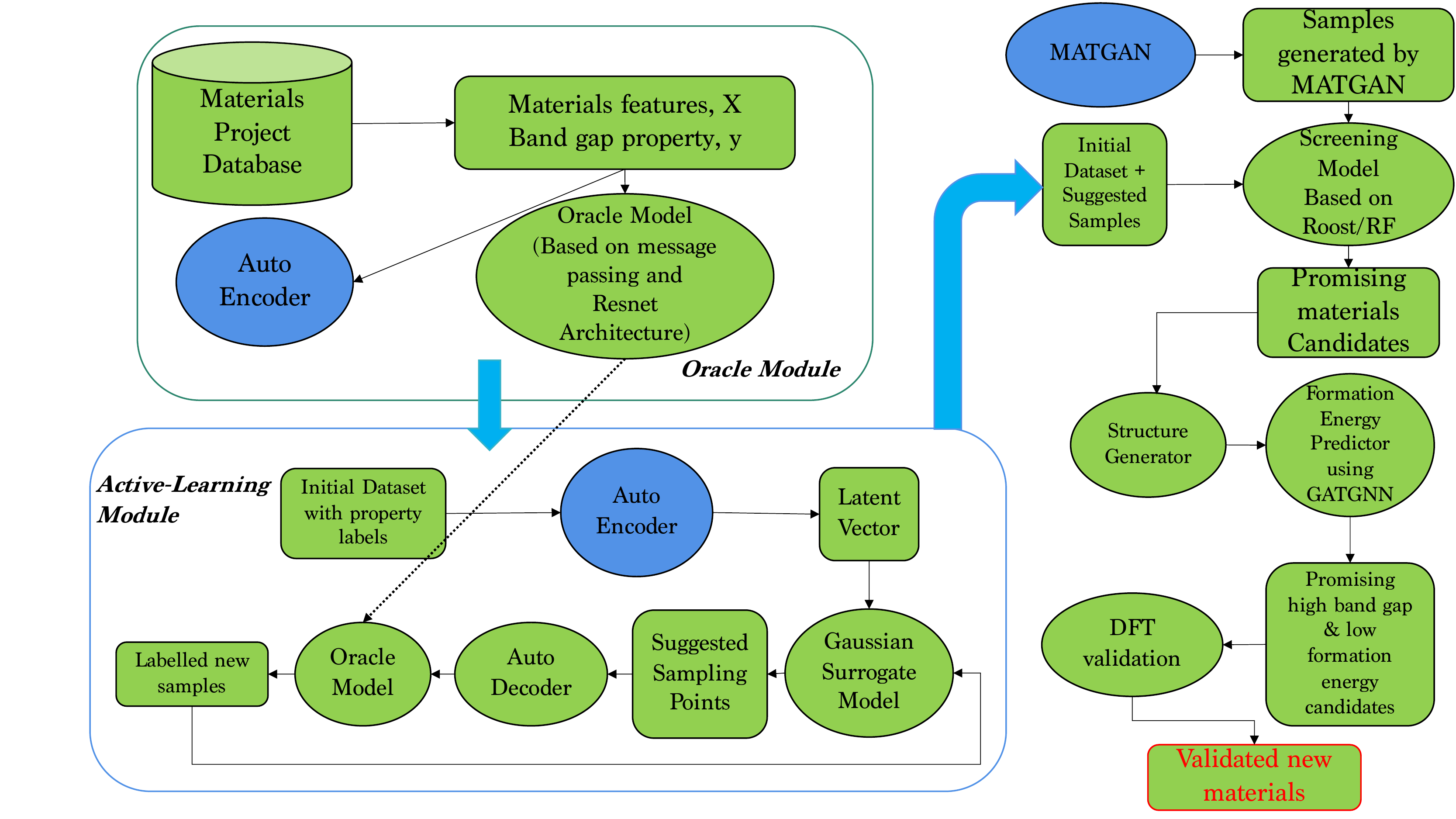}
  \caption{Framework for active generative design for materials discovery.}
\end{figure}

\textbf{Oracle Module}

The Oracle Module is a machine learning model trained with material features and band gap labels of the whole dataset in our experiments. This module aims to simulate the DFT function and facilitate the prediction of band gap property based on a vector of material features. In practice, this Oracle model should be replaced with DFT calculations or experiments.  The descriptors for our Oracle model are calculated using Roost (Representation Learning from Stoichiometry) \cite{goodall2019predicting}. Roost is a machine learning algorithm that learns the stoichiometry-to-descriptor map directly from data, the key idea of which is each material’s composition is represented as a dense weighted graph.
Roost uses Matscholar embedding \cite{tshitoyan2019unsupervised} for embedding and CGCNN \cite{xie2018crystal} for internal representation. The output network used for the reference model is a Resnet with five hidden layers and ReLU activation functions.

\textbf{Active Learning Module}

The function of the Active learning Module is to actively explore the whole design space to identify the informative sampling points and obtain their property annotations, which are fed to train a highly accurate property prediction model. 
In previous active learning-based design studies such as \cite{min2020accelerated}, the samples are directly represented as 145 common chemical features, which are used to train the Gaussian Surrogate model. New sampling points are not generated but selected from existing entries in a given database. However, such physical feature representation of materials is not applicable for active learning in the uncharted design space because the sampled feature representations can not be easily mapped back to valid materials formulas/structures. Actually, this reverse mapping procedure usually makes it problematic for the Gaussian Surrogate model to generate meaningful samples that can be decoded into chemical formulas. To address this representation issue, we introduce the AutoEncoder deep neural network model as a latent vector generator. It can be used to convert a chemical formula into a 128-dimension latent vector and vice versa. Thus, formulas are not only well represented but also provide latent vector boundary for each dimension which makes it easier for the optimization process. To get the property annotation by the Oracle model, the recommended sampling points from the Surrogate model will be first decoded into a formula, and then its Magpie descriptor features are calculated, and the band gap property will be predicted by the Oracle model. 

Latent vectors as generated above, along with their property annotations, will be fed as input to the Gaussian Surrogate model, which will give uncertainty-based sampling point suggestions for Oracle Module to evaluate. This suggested vector has to be converted back to a chemical formula in order to get material features. However, cases might occur when some suggested latent vectors can not be decoded correctly into formulas and will negatively affect the design space exploration performance of the Active-Learning module.
Afterward, labeled new samples will be sent back to the Gaussian Surrogate model for a new round of recommendations.
It should be noted that different initial datasets and the Gaussian Surrogate model constructed both have an influence on the sampling points, which affect the final performance of the ML screening model trained with these data points.


\subsection{Autoencoder neural network for latent space sampling}

\begin{figure}[h]
  \centering
  \includegraphics[width=0.95\linewidth]{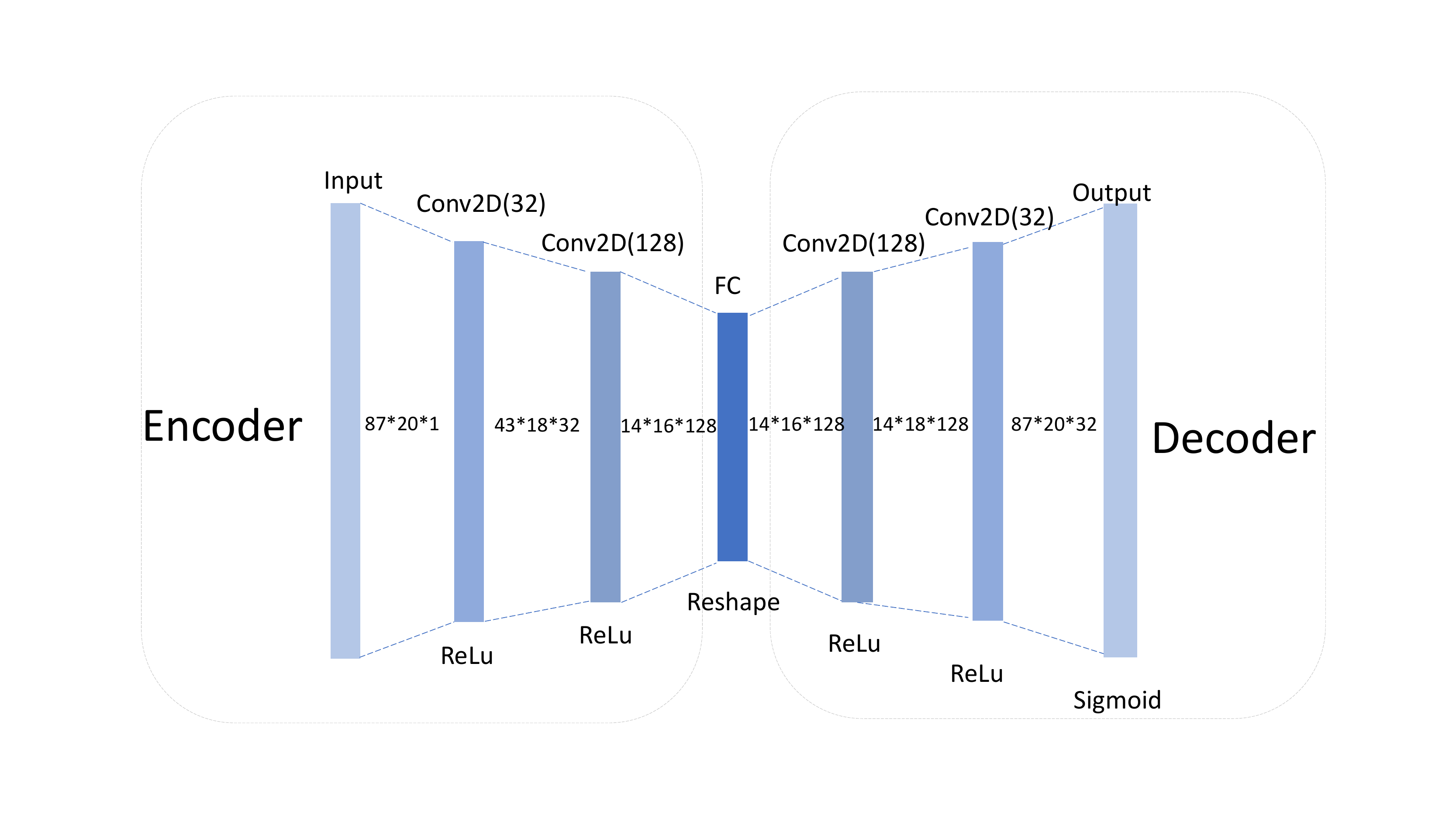}
  \caption{Autoencoder model.}
  \label{fig:Autoencoder}
\end{figure}

Standard autoencoder is a popular representation learning and dimension reduction method, which is composed of two parts: an encoder and a decoder. Our autoencoder model is used to map materials compositions into a continuous space, which can be used by the Bayesian optimizer to sample the design space. The sampled points can then be decoded back into materials compositions. In our model, the encoder has 
two convolutional layers followed by a fully connected layer and a decoder consists of a fully connected layer and three convolution layers. The autoencoder (AE) model is shown in Fig \ref{fig:Autoencoder}. The ReLU is used as activation function in each neural layer except the last layer of the decoder for which the Sigmoid function is used instead. We use the Adam optimizer to train the model parameters and the learning rate is set as $0.001$. We use the cross-entropy loss for the training. The batch size is set as 128. The autoencoder is trained with 66324 inorganic materials selected from the Materials Project database. 

\begin{figure}[h]
  \centering
  \includegraphics[width=0.95\linewidth]{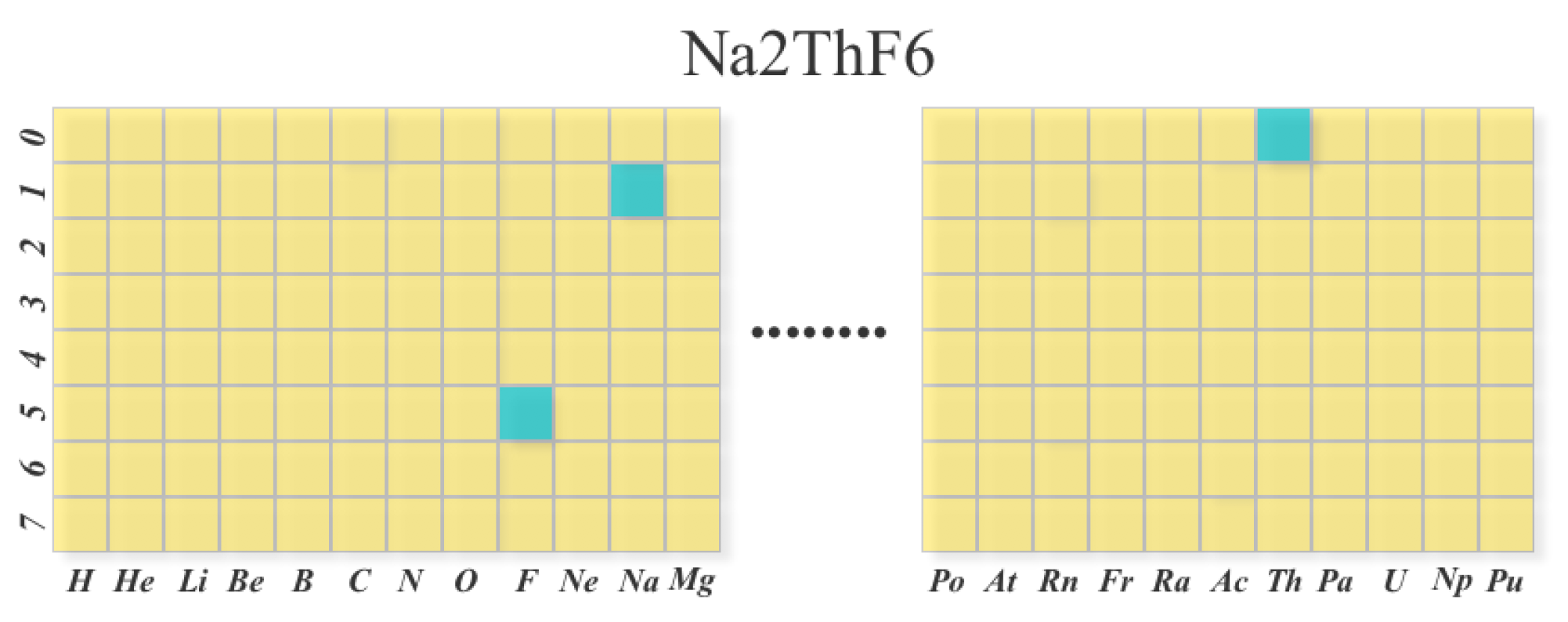}
  \caption{One hot encoding of material compositions}
  \label{fig:onehot}
\end{figure}

The material compositions are represented using the one-hot encoding. Through the simple statistical calculation of the materials in the MP dataset, 87 elements are found, and each element usually has less than 20 atoms in any specific compound/formula. We then represent each material as a sparse matrix with 0/1 cell values. Each column represents one of the 87 elements, while the column vector is a one-hot encoding of the number of atoms of that specific element. The whole encoding scheme is shown in Figure \ref{fig:onehot}.

\subsection{Active learning module for efficient sampling}

The active learning module in our pipeline is used for efficient exploration of the design space to identify informative samples and obtain their property annotations, which allows generating more training samples to train an accurate property prediction model. The sampled materials can also be used for screening candidates from them with desirable properties. We use Bayesian Optimization (BO) \cite{snoek2012practical} as the active Learning sampling strategy to explore the huge design space. 

BO is an Efficient Global Optimization (EGO) algorithm aiming to find the maximum value of a costly unknown function in as few iterations as possible. BO builds Gaussian Process Regression, a surrogate model, to not only evaluate the objective function but also quantify the uncertainty with each prediction.  Acquisition functions such as Expected Improvement (EI) \cite{mockus1978application} and Upper Confidence Bound (UCB)\cite{srinivas2009gaussian} are conducted to find the next point to evaluate. 

\begin{equation}
\mathrm{EI}(\mathbf{x})=\left\{\begin{array}{ll}
\left(\mu(\mathbf{x})-f\left(\mathbf{x}^{+}\right)-\xi\right) \Phi(Z)+\sigma(\mathbf{x}) \phi(Z) & \text { if } \sigma(\mathbf{x})>0 \\
0 & \text { if } \sigma(\mathbf{x})=0
\end{array}\right.
\label{EI}
\end{equation}

Where $\mu(\mathbf{x}$)  \mbox{ and } $\sigma$($\mathbf{x}$)  \mbox{ are the mean and standard deviation of GP posterior at point x. } $\Phi$ \mbox{  and } $\phi$  \mbox{ are the Cumulative }

Distribution Function (CDF) and Probability Density Function (PDF) of the standard normal distribution, respectively whereas $Z$ \mbox{is denoted in Equation \ref{Z}.} $f\left(\mathbf{x}^{+}\right)$ \mbox{represents the best observed value} whereas x is the corresponding location.

The first term in the upper Equation \ref{EI} denotes exploitation, and the second summation term is the exploration term. EI can adjust the level of balance between exploitation and exploration through changing parameter $\xi$. 
The higher $\xi$ is, the level of importance of posterior mean value improvement will decrease, while the algorithm will be more inclined to explore unknown areas.

\begin{equation}
Z=\left\{\begin{array}{ll}
\frac{\mu(\mathbf{x})-f\left(\mathbf{x}^{+}\right)-\xi}{\sigma(\mathbf{x})} & \text { if } \sigma(\mathbf{x})>0 \\
0 & \text { if } \sigma(\mathbf{x})=0
\end{array}\right.
\label{Z}
\end{equation}

Our framework adopted Acquisition function UCB in Equation \ref{UCB_equation} to probe promising points. Intuitively, UCB explores the design space to locate the max value by considering both expected performance and uncertainty across all solutions. $\kappa$ factor here enables fine adjustments between exploration and exploitation.
\begin{equation}
\mathbf{x}_{t+1}=\underset{\mathbf{x}}{\arg \max }\left(\mu_{t}(\mathbf{x})+\kappa \sigma_{t}(\mathbf{x})\right)
\label{UCB_equation}
\end{equation}


Fernando Nogueira's package \cite{nogueira2014bayesian} is used as the sampler to provide suggestions about which latent vectors will be generated and probed during the search process. Depending on the experiments, a different number of initial sampling points (prior) are used to build the Gaussian Process surrogate model. In every iteration, the sampler will suggest 20 sampling points. Only those candidates the can be decoded into a valid material composition will be registered and used to update the Gaussian Process model. The sampler will stop sampling when the setting budget is reached.

\subsection{Template based structure prediction and DFT based validation}

To validate the band gap property of promising candidates generated by our framework, we perform a template-based structure prediction over the generated compositions to create their structures for DFT calculation after charge neutrality and electronegativity integrity checking. 

We first measure the similarity between a candidate chemical formula and all the formulas from the Materials Project(MP) database using Earth Mover’s Distance (EMD) \cite{hargreaves2020earth}. EMD computes the ratio of each of the elements and the absolute distance between the elements on the modified Pettifor scale between two compositions, which leads to greater alignment with chemical understanding than the Euclidean distance. The MP formulas are then ranked by their EMD distance, and the CIF file of the most similar formula is identified as the structure template for element substitution to generate the structures for the candidate compositions. The CIF files of these new structures will then be used as input for a pre-trained Global Attention Graph Neural Network (GTGNN) \cite{louis2020graph} to predict their formation energy. Only candidates with negative formation energy will be kept. Finally, DFT calculations will be performed over these filtered structures to calculate their band gaps and formation energy. Phonon calculation is also applied to one of the discovered materials to check its thermodynamic stability.

\section{Experiments}

\subsection{Band gap Dataset}

\begin{figure}
  \centering
  \includegraphics[width=0.6\linewidth]{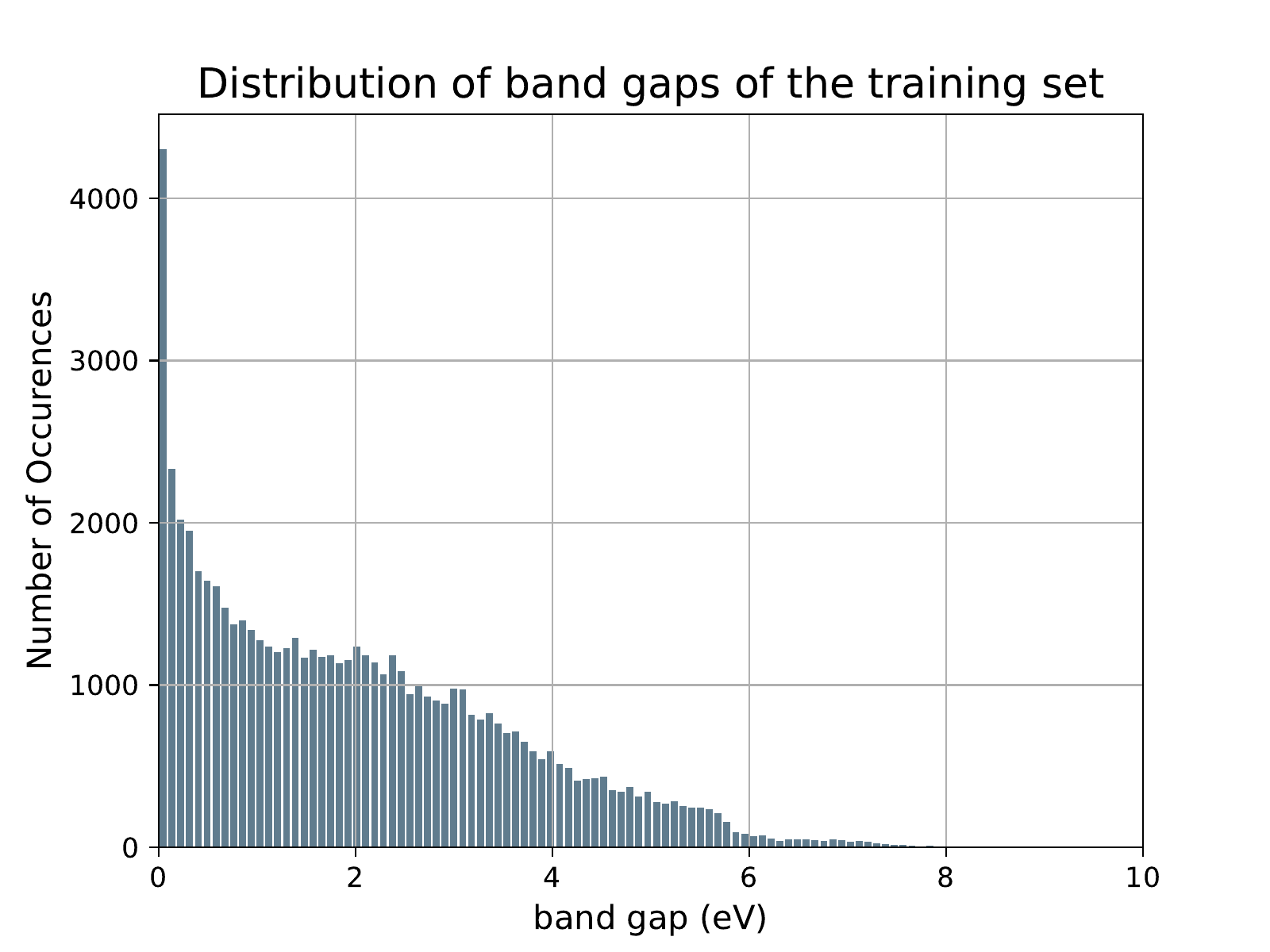}
  \caption{Distribution of band gaps of the training set.}
\label{fig:distribution_of_bandgap_dataset}
\end{figure}

We download the Materials Project dataset from http://materialsproject.org and remove duplicate composition entries to get 42,667 samples with band gap data. We call this the MP dataset. 
The distribution of band gaps of the MP dataset is shown in Figure \ref{fig:distribution_of_bandgap_dataset}.

\subsection{Experimental Design}
To validate the performance of our active generative design framework, 1,180,000 potential material compositions are generated via the MATGAN network \cite{dan2020generative}. Bayesian Optimization with the UCB utility function is used as active learning to explore promising sampling points for machine learning model training. $Kappa$ is set to 100 since the goal is to generate diverse samples. The Roost band gap prediction model is trained based on the original and extra generated sampling points so as to screen high band gap materials out of these potential materials.
In order to better evaluate the effectiveness of our AGD framework, we designed four experiments according to the availability of the different amounts of training data. Each model is evaluated by repeating the experiment 10 times to report statistically significant results. In each experiment, we split the whole dataset into 60\% for training, 20\% for validation, and 20\% for testing, respectively.

1) Exp1-oracle model: 42,667 materials from the Materials Project database are used for training the Oracle model for band gap prediction, which is then used to screen high potential band gap materials from MATGAN-generated candidates. This is the ideal case where there are abundant annotated materials samples for training an ML property prediction model for screening.

2) Exp2: We randomly select 1,000 labeled samples from the MP dataset to train a baseline ML model (Exp2 BS). We then use the Gaussian Surrogate model for active learning. After 1,000 sampling points are suggested by the Bayesian optimizer, 741 samples can be decoded into valid compositions. We use these newly-explored samples together with the 1000 initial samples to update our training dataset and train an updated Screening Model (Exp2 AL). Both band gap prediction models will be evaluated and used for screening MATGAN-generated candidates.

3) Exp3: Only 300 samples are randomly selected from the MP dataset for the baseline ML model (Exp3 BS) training and Gaussian Process initialization. We then generate 1,000 new samples using the active learning model, which leads to 853 valid compositions. In total, 1153 samples are used to train the updated ML model (Exp3 AL). We compare the performance between the baseline screening model (BS) and the auxiliary Active Learning trained screening model (AL) as we do in Exp 2.

4) Exp4: Only 300 samples whose band gap values are below 3.0 eV are selected for active learning. This control experiment aims at validating the performance of our AGD framework when only small, poor quality dataset is available.

To compare the performance of the model performances trained with a different number of samples, as shown above, we pick the 8535 test samples from the Exp1 and remove the entries that appear in the Exp2 and Exp3 experiment datasets. We then select 500 samples from the remaining samples as the test set for evaluating all the models.

Additionally, a control experiment is conducted in which we adjust the parameter kappa of the utility function UCB to inspect the active learning behavior between exploration and exploitation.

\subsection{Results}

\subsubsection{Prediction Model Capability with Active Learning}
Table \ref{table:MP} shows model performance and the number of candidate materials with a band gap greater than 6.0 eV found with or without active learning. Exp 1 has the most available training data, which results in an average $R^2$ performance of 0.735 when evaluated on its 20\% hold-out test set.  The baseline models trained with 1,000 and 300 samples in Exp 2 and Exp 3 have corresponding $R^2$ scores as low as 0.361 and -0.243, showing the under-performance due to lack of sufficient training samples. The negative $R^2$ score of the baseline model performance of Exp3 BS shows that 300 initial samples are not sufficient to train a reliable model. In contrast, 1,000 initial samples can train a much better model. However, after we apply the active learning process, the great performance gain is achieved as the model metric $R^2$ scores are improved from 0.361 to 0.611 in Exp2 and from -0.243 to 0.757 in Exp3 when evaluated using the random hold-out test over the dataset composed of initial samples and newly sampled points. However, the improvements are a little bit over-estimated here as we can observe that the performance improvement when evaluated with the external 500-sample test set, the improvement due to active learning is reduced to 0.19 for Exp3 AL and is only 0.005 for Exp2 AL. After close examination, we find that this is due to the mismatch of the training set distribution and the external test set distribution: most of the samples suggested by active learning during the search process have high band-gap values greater than 2 eV while the majority of the whole dataset from which the 500-sample test set are drawn from have band gap values between 0 and 2 eV (See Figure 4). This shows the samples generated during the active-learning search are highly biased toward candidates with higher band gaps.

During active learning of Exp2 and Exp3, we find that 853 decodable new materials out of 1,000 Bayesian Optimization iterations are found and evaluated in Exp3 AL, while 741 new materials are found in Exp2 AL. The reason that Exp2 AL finds fewer new materials given the same budget is that more initial samples will lead to a better Gaussian Process model with less uncertainty so that the optimizer tends to suggest smaller steps on the latent vector. Such latent vectors with small changes tend to result in decoding into the same materials.

\begin{table}[h!]
\centering
\caption{Average Model performance and screening results}
 \begin{tabular}{l c c c c c c c} 
 \hline
 Exp No. & \#samples & $R^2$(hold out) & $R^2$ (test set) & MAE & RMSE & \#candidates & Recovery Rate\\ 
 \hline
 Exp1 & 42,677 & \textbf{0.735$\pm$0.021} & \textbf{0.850$\pm$0.023} & \textbf{0.505$\pm$0.016} & \textbf{0.826$\pm$0.034} &102 & 100\%\\ 
 Exp2 BS & 1,000 & 0.361$\pm$0.022 & 0.409$\pm$0.016 & 0.882$\pm$0.024 & 1.208$\pm$0.021  & 123 & 51.96\%\\ 
 Exp2 AL & 1,741 & \textbf{0.611$\pm$0.009} & 0.414$\pm$0.013 & \textbf{0.539$\pm$0.008} & \textbf{0.869$\pm$0.010} & \textbf{171} & \textbf{78.43\%}\\
 Exp3 BS & 300 & -0.243$\pm$0.082 & 0.227$\pm$0.042 & 1.202$\pm$0.057 & 1.633$\pm$0.054 & 33 & 2.94\%\\ 
 Exp3 AL & 1,153 & \textbf{0.757$\pm$0.015} & \textbf{0.417$\pm$0.005} & \textbf{0.309$\pm$0.019} & \textbf{0.592$\pm$0.019} & \textbf{81} & \textbf{57.84\%}\\ 
 \hline
\end{tabular}

\label{table:MP}
\end{table}

In Exp4 AL, we perform active learning based on randomly sampled 300 materials whose band gap is below 3.0 eV. Figure \ref{UCB} shows our framework can easily find high band gap potential materials (>6.0 eV) within 30 iterations. However, these found materials may not be valid if chemical rules are applied. As a result, most of these generated labeled samples will function as data augmentation measurements for screening model training.

\subsubsection{Exploration versus exploitation in active learning: control experiments}

We conduct several control experiments to find the relationship between exploration and exploitation during the active learning process. To find the suitable trade-off between them, we conduct experiments using the Upper Boundary Confidence (UCB) acquisition function, for which the Kappa parameter can be used to control the exploration rate. Larger Kappa leads to more exploration, while smaller one leads to more exploitation.

\begin{figure}[h]
  \centering
  \begin{subfigure}{.45\textwidth}
      \includegraphics[width=1.1\linewidth]{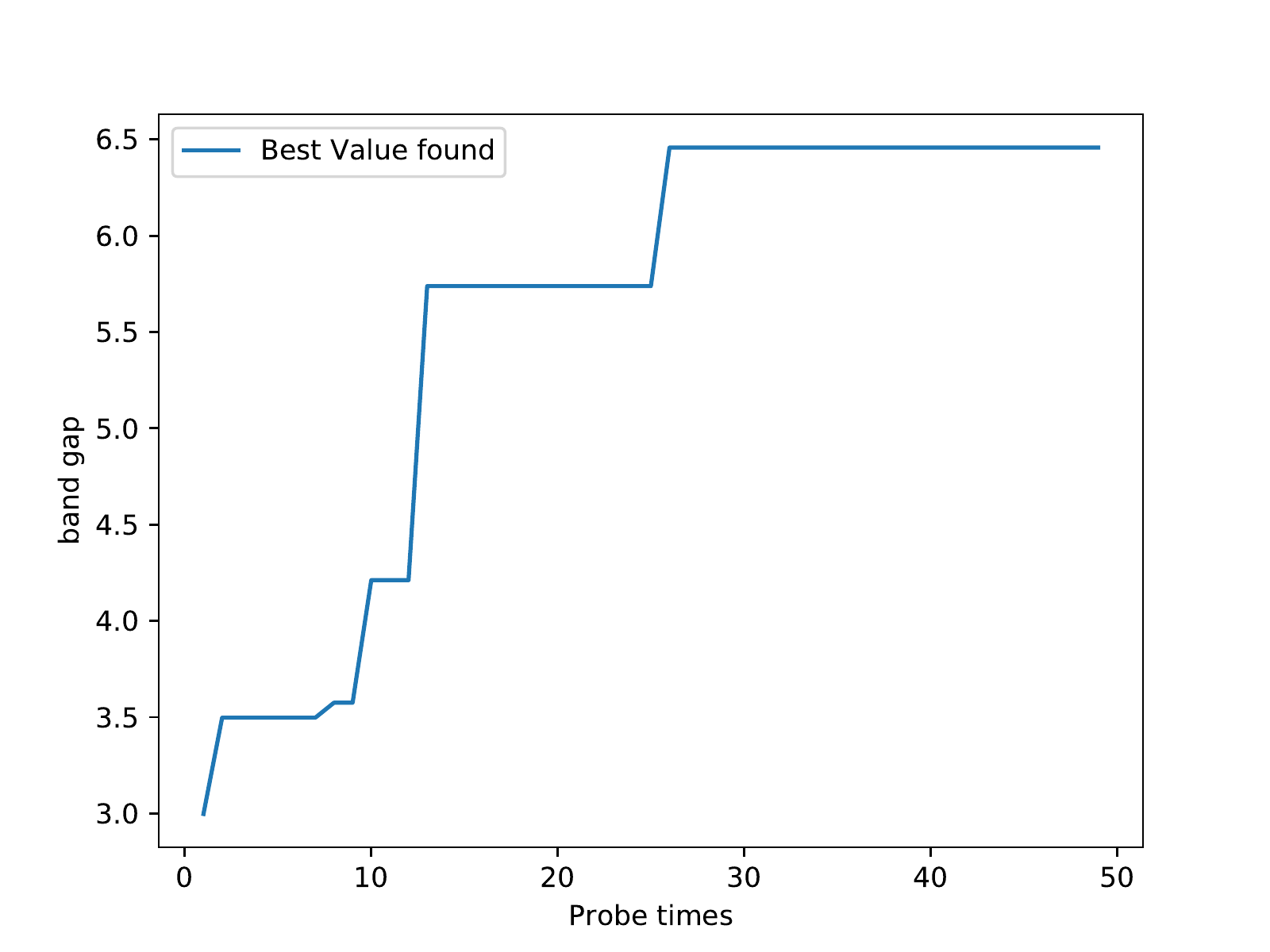}
      \caption{Band gap trace of the sampled materials during the search process}
      \label{UCB}
      \vspace{1pt}
  \end{subfigure}
  \begin{subfigure}{.45\textwidth}
      \includegraphics[width=1.1\linewidth]{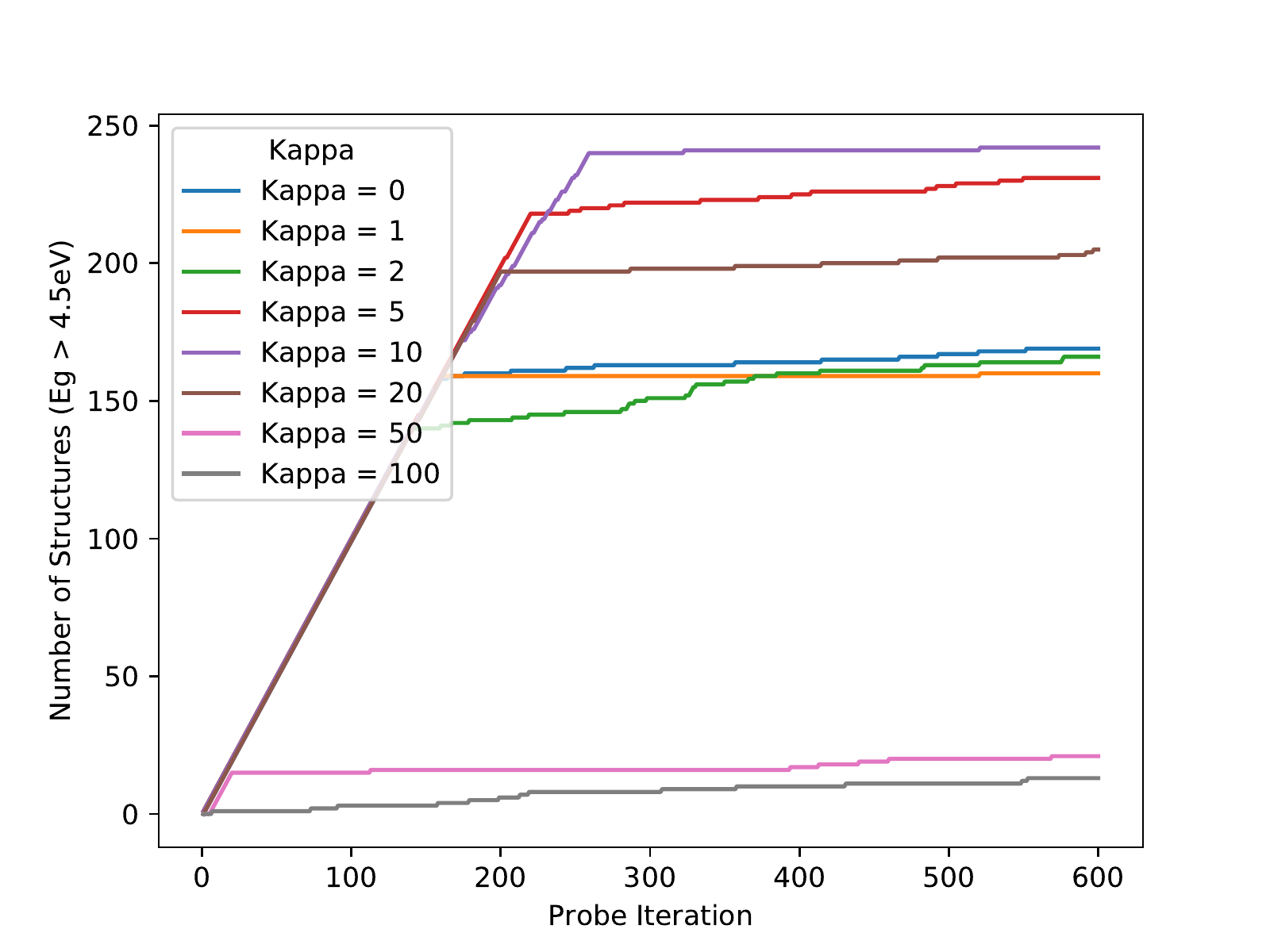}
      \caption{Exploration vs Exploitation (UCB utility function).}
    \vspace{11pt}
  \end{subfigure}
  \label{fig:experiment boxplot}
  \caption{Active Inverse Design Process}
\end{figure}

The relationship between exploration and exploitation is shown in Figure \ref{UCB}. The blue line represents pure exploitation, which flattens after about 150 iterations since nearly all candidates with high expected values have been utilized without exploration.
As Kappa gradually increases, which means UCB explores more, the algorithm will more likely jump out of local minima. As a result, the number of identified high band gap structures almost always increases when we increase Kappa from 0 to 10. 
However, the number of found structures decreases when Kappa increases from 10 to 100 as active learning continues to lean towards exploration. 
The grey line (Kappa=100) and pink line (Kappa=50) become flatten almost from the beginning since uncertainty has a rather large enhancement as shown in formula \ref{UCB_equation}.
As a result, active learning arrives at its trade-off and maximizes the discovery of promising structures when we set Kappa to 10.

\subsubsection{Finding new materials with targeted band gaps using the active generative design framework}

To understand the effect of using active learning, we apply the trained ML models to screen the MATGAN-generated hypothetical materials and inspect their true band gaps calculated using the Oracle model .

The last two columns of Table 1 shows the number of candidate compositions with predicted band gaps greater than 6 eV and the accuracy measured by the overlap percentage of candidates with respect to the candidates screened out by the Oracle model (Exp 1).  The recovery rate column of Table 1 also shows models trained with active learning (AL) augmented data achieve a much higher recovery rate than the baseline model (BS).  Specifically, with 1,000 search budget, Exp2 AL and Exp3 AL improved the recovery rates of the baseline models by 26.47\% and 54.90\% respectively, indicating that active learning can improve the screening capability of high band gap materials with the ML models trained with suggested samples and initial samples although they $R^2$ score as evaluated by the external dataset is improved just a little.

\begin{figure}[h]
  \centering
  \begin{subfigure}{.45\textwidth}
    \includegraphics[width=\textwidth]{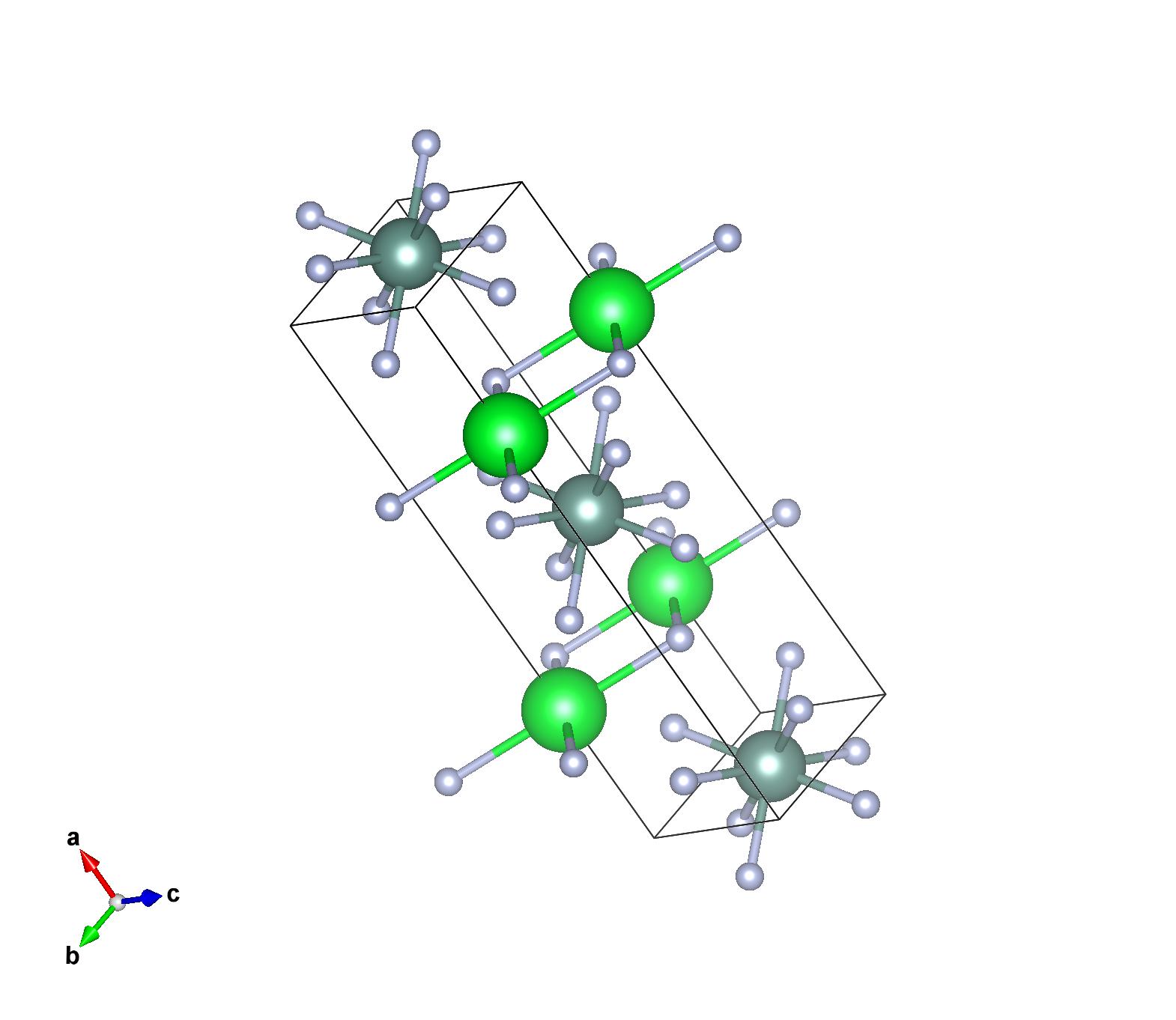}
    \caption{Structure of SrYF$_5$}
    \vspace{3pt}
  \end{subfigure}
  \begin{subfigure}{.45\textwidth}
    \includegraphics[width=\textwidth]{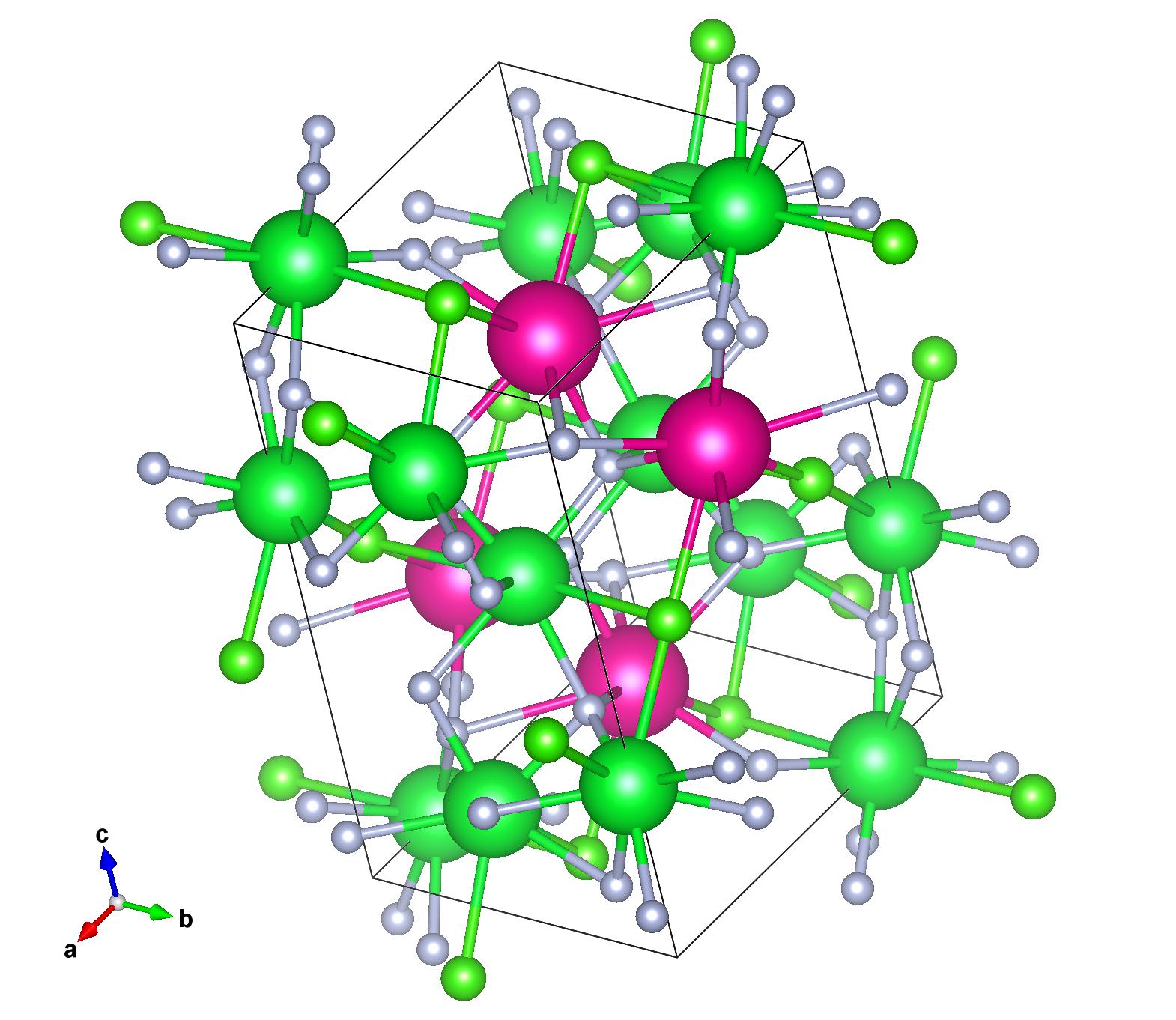}
    \caption{Structure of RbSr$_2$ClF$_4$}
    \vspace{3pt}
  \end{subfigure}

  \caption{Structures of the discovered new materials.}
  \label{fig:structures_found}
\end{figure}

From the screening sets, we have found a high band gap material SrYF$_5$ in Exp1, Exp2 AL, and Exp3 AL whose predicted band gap is 6.71, 7.02, 7.11eV, respectively. We use the element substitution to get its structure (Figure\ref{fig:structures_found}(a))and calculate its band gap using DFT calculation, which shows a band gap value of 6.42eV. However, SrYF$_5$ does not appear in the screening results of models from Exp2 BS and Exp3 BS, which are the baseline models constructed without conducting active learning. This demonstrates the benefit of active learning in new materials discovery.

Additionally, our active learning process has found a new material RbSr$_2$ClF$_4$ (structure shown in Figure\ref{fig:structures_found}(b)) directly during the active learning process. Its DFT validated band gap value is 5.64eV. This material is a minor variation of the material RbSr$_2$Cl$_5$ (band gap: 5.20eV), a member of the initial dataset for active learning training. Altogether, the results show that our active generative design framework is capable of working well when only small training dataset is available by efficiently exploring the design space and improving ML model performance.

\subsubsection{Finding new materials with MATGAN+screening approach}

The main idea of active generative design is to use active learning to train a good screening property prediction model, starting with limited training samples. When there is a sufficient number of training samples, one can simply build an ML model as the filter to screen the hypothetical samples generated by GAN models. When there is limited labeled training samples, it is better to use active learning to improve the ML model using as few expensive (e.g., DFT) sample annotation as possible. To prove the utility of active learning for such screening-based new materials discovery, we applied it to the high band gap materials screening problem starting with only 300/1000 labeled samples.

We first use the MATGAN, a generative adversarial network (GAN) for efficient generation of new hypothetical inorganic materials \cite{dan2020generative}. Trained with materials from the ICSD database, the MATGAN model generates 2.99 million inorganic formulas, most of which are charge-neutral and electronegativity-balanced. To simplify the screening process, we trim the generated formula set by filtering out those formulas with the total atom numbers more than 20 or the number of elements greater than five, or the number of atoms for each element more than 7. After checking chemical validity, the total number of candidates for screening is 1.18 million, and we call these samples the MATGAN dataset. We then use the whole MP dataset to train a Random Forest model to screen this dataset for potential semiconductors with a band gap between 2.0 eV and 4.0 eV. Similarly, we use the initial samples in Exp1, and Exp2 plus the active learning explored samples to train two Roost ML models and use them to screen the MATGAN dataset for band gaps greater than 6.0 eV. From this final candidate list, we select 18 for DFT formation energy and band gap validation.

\paragraph{DFT based Calculations}

The density functional theory (DFT) based first principle calculations were carried out based on the  Vienna \textit{ab initio} simulation package (VASP) \cite{Vasp1,Vasp2,Vasp3,Vasp4}.  The projected augmented wave (PAW) pseudopotentials were considered for the electron-ion interactions with 400 eV plane-wave cutoff energy \cite{PAW1, PAW2}.  The exchange-correlation functional was treated by using the generalized gradient approximation (GGA) based on the Perdew-Burke-Ernzerhof (PBE) method \cite{GGA1, GGA2}. The energy convergence criterion was set as 10$^{-5}$ eV. The atomic positions were optimized until the forces become less than 10$^{-2}$ eV/{\AA}. The Brillouin zone integration for the unit cells was performed using the $\Gamma$-centered  Monkhorst-Pack $k$-meshes. The Formation energies (in eV/atom) of several materials were calculated based on  Eq.~\ref{eq:form}, where $E[\mathrm{Material}]$ is the total energy per unit formula of corresponding material, $E[\textrm{A}_i]$ indicates the energy of $i^\mathrm{th}$ element of the material, $x_i$ represents the number of A$_i$ atoms in a unit formula, and N is the total number of atoms in a unit formula($N=\sum_i x_i$). Those values are reported in Table~\ref{table:DFT}. CaClF$_5$ (-2.09 eV/atom), SrYF$_5$ (-4.10 eV/atom), and SrClF$_3$ (-2.55 eV/atom) show very low formation energies, implying our model is capable of finding new semiconductor materials, which are thermodynamically highly stable relative to the parent compounds of their elements.

\begin{equation}
    E_{\mathrm{form}} =\frac{1}{N}(E[\mathrm{Material}] - \sum_i x_i E[\textrm{A}_i])
    \label{eq:form}
\end{equation}

\begin{table}[h]\centering
\caption{DFT verified formation energies and the band gaps of several semiconductors}
\label{table:DFT}
\begin{tabular}{cccc}
\hline
 Material& $E_\mathrm{form}$(eV/atom) & $E_\mathrm{gap}$ (eV) & Method   \\
 \hline 

 SrYF$_5$&  -4.10  &  6.42 & AL+Roost with MATGAN\\
 RbSr$_2$ClF$_4$ & -3.35 & 5.64 & Direct Active Learning Search\\
 SiO$_2$F& -1.69  &3.99 & Direct Active Learning Search\\
 PO$_2$F & -2.10  &3.84 & Direct Active Learning Search\\
 SH$_2$O$_3$ & -0.10  &3.74 & Direct Active Learning Search\\
 SiF$_3$ & -4.27  &3.79 & Direct Active Learning Search\\
 NaF$_5$C & -1.78  &4.80 &AL+RF with MATGAN\\
 CaClF$_5$  &-2.09 &3.46 &AL+RF with MATGAN\\
 YCl$_3$ &  -1.94  &2.18 & AL+RF with MATGAN\\
 SrClF$_3$ &  -2.55  &3.18 & AL+RF with MATGAN\\

 SrCl$_2$F$_3$ &  -1.91 &2.27 & AL+RF with MATGAN\\
 AlSeCl & -0.58 &3.70 & AL+RF with MATGAN\\
 As$_2$SO$_3$ & -1.16  &2.72 & AL+RF with MATGAN\\
 
 \hline
\end{tabular}

\end{table}

\begin{figure}[h]
  \centering
  \begin{subfigure}[b]{.45\textwidth}
    \includegraphics[scale=0.3]{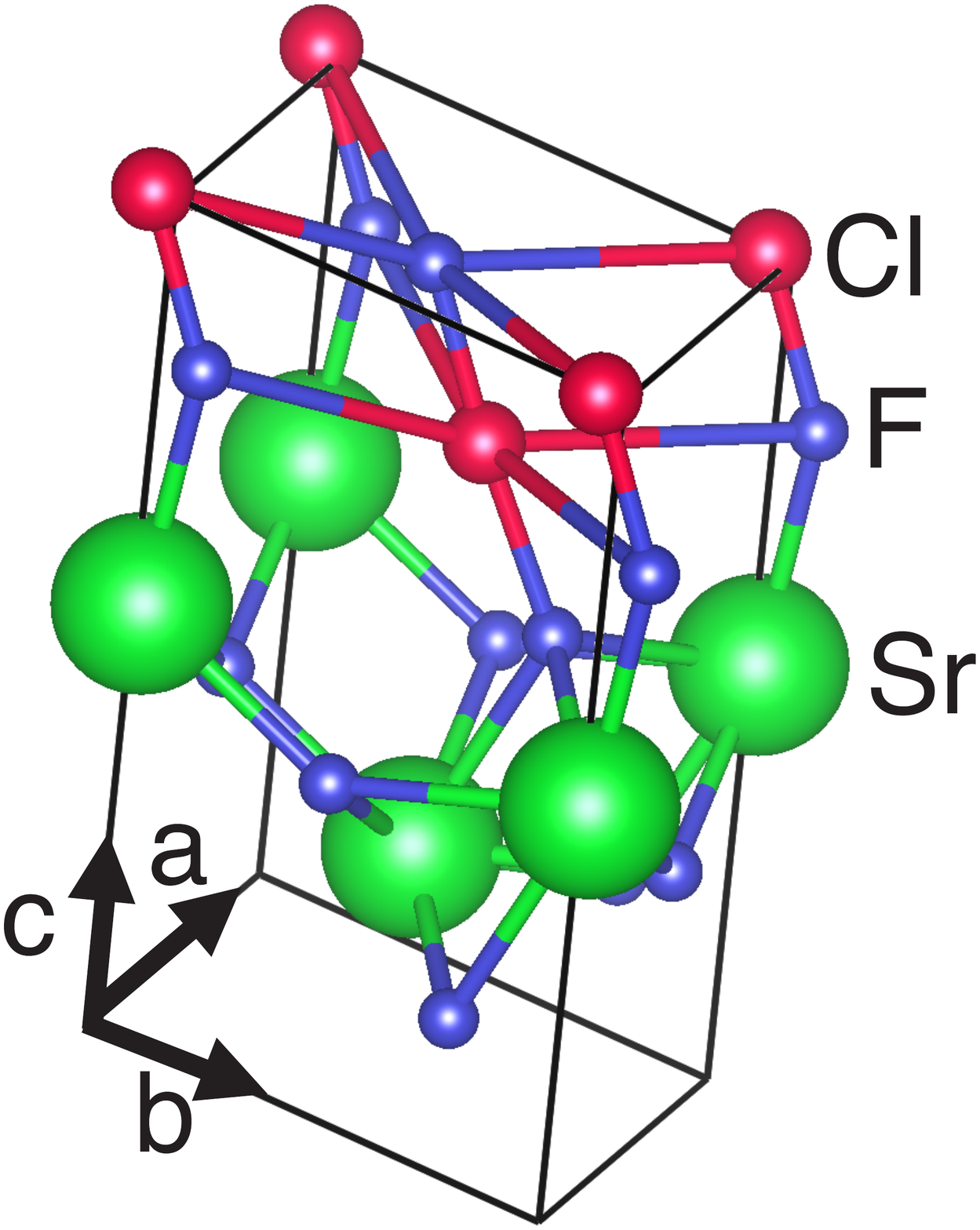}
    \caption{The structure of SrClF$_3$}
    \vspace{2pt}
  \end{subfigure}
  \begin{subfigure}[b]{.45\textwidth}
    \includegraphics[scale=0.35]{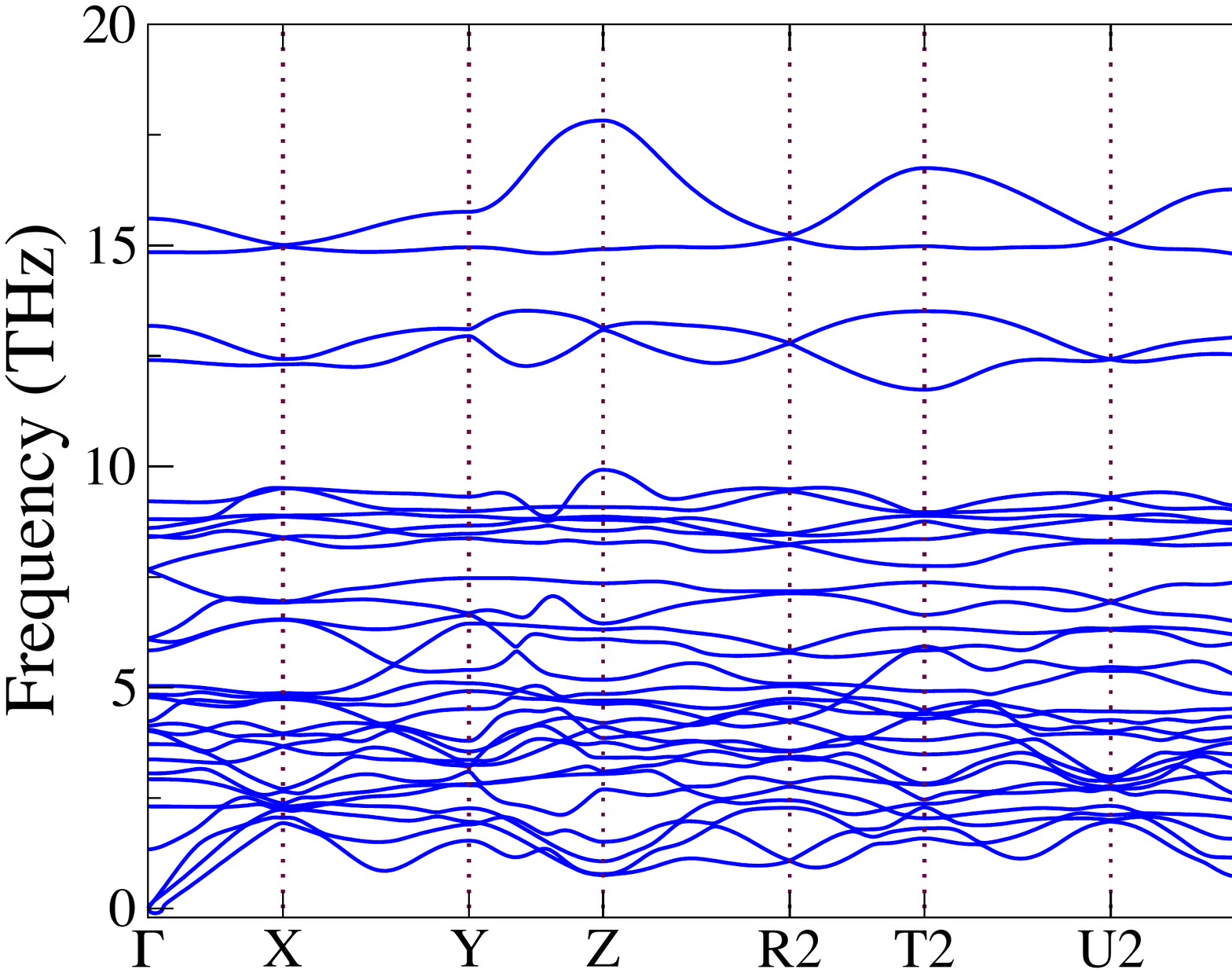}
    \caption{Phonon dispersion curves for SrClF$_3$ }
    \vspace{3pt}
  \end{subfigure}

  \caption{Properties of the discovered new materials.}
  \label{fig:phonon}
\end{figure}

Moreover, we find that SrClF$_3$ is a stable semiconductor with P1 space-group symmetry (Figure\ref{fig:phonon}). The unit cell lengths of the material are $a= 4.44$  {\AA}, $b= 4.75$ {\AA} and $c=8.30$ {\AA}, while the unit cell angles are $\alpha= 102.3$, $\beta= 90.1$ and $\gamma= 90.1$.  The positive frequencies in the phonon calculations (see Figure \ref{fig:phonon} (b)) show that the material is dynamically stable at 0K temperature. The energy above hull, which is 0.046 eV/atom, is obtained against the competing phases SrF$_2$ and ClF using Pymatgen code \cite{Pymatgen}. 


We also carry out density functional perturbation theory (DFPT) calculations as implemented in VASP code for studying the mechanical properties of SrClF$_3$. Due to the P1 space-group symmetry of this material, there are 21 independent elastic constants. The calculated $C_{11} = 64.19$ GPa and $C_{12}= 62.87$ GPa elastic constants are almost equal because of $a \approx b$, and  $\alpha =\beta$ lattice parameter relationships. Bulk modulus (41.26 GPa), Shear modulus (15.61 GPa), and Young's modulus (41.587) were calculated based on Hill approach \cite{Hill_1952}. When an infinitesimal strain ($\epsilon$) is applied on a material, the total energy of that material can be written as $E=E_0 + \frac{1}{2}V_0\sum_{i,j=1}^{6}C_{i,j}\epsilon_i\epsilon_j + O(\epsilon^3)$. Here, C represents the matrix of second-order elastic constants. The Born elastic stability conditions show that the matrix should be definite positive, all eigenvalues of $C$ should be positive, and all the principal components should be positive. Even though most crystal symmetries have simple relationships between the elastic constants to show whether the materials are mechanically stable, triclinic systems have very complicated equations because of having 21 independent elastic constants \cite{Born_Criteria}. Thus, we used VASPKIT \cite{VASPKIT} code to analyze the elastic properties, and it confirms that SrClF$_3$ is very likely to be a mechanically stable semiconductor.

\section{Discussion}

\begin{figure}[h]
  \centering
  \includegraphics[width=1.2\linewidth]{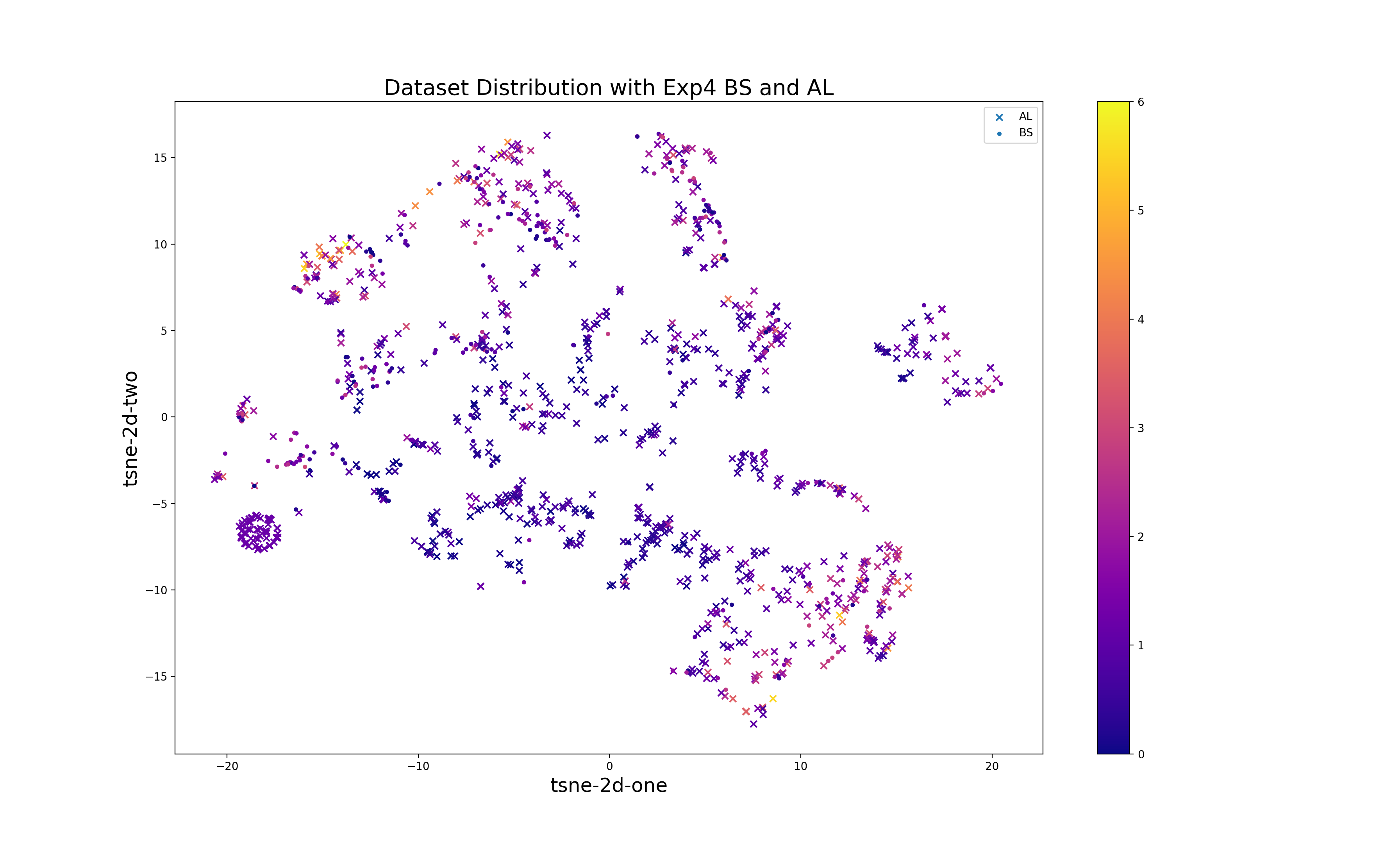}
  \caption{Comparison of the sample distributions of the base line training set and the AL enhanced dataset using t-sne visualization. It is found that the samples traversed by the active learning algorithm are much more diverse and span the majority of the design space.}
\label{fig:distribution}
\end{figure}

Our active learning experiments have shown that it can find new materials with high band gap values as validated by DFT, demonstrating that active learning-based generative design has the capability to discover new material with target properties using guided/directed search. Figure \ref{fig:distribution} shows how active learning explores the inorganic material space. Active-learning probed samples ( circles) are obviously more widely distributed than the original training set (+). However, there are several aspects to improve for our active generative design framework.

First, currently, we are using Bayesian Optimization as the Efficient Global Optimization (EGO) strategy to search the 128-dimension design space containing the entire inorganic materials. Yet, many latent representations sampled during active learning search cannot be decoded into valid chemical formulas, which has caused a significant waste of the computational budget and decreases the efficiency of the Bayesian Optimization sampler. Additional physicochemical constraints need to be incorporated into the sampling strategy so as to achieve more efficient sampling. Second, we encountered the out-of-distribution prediction issue of standard machine learning: our random forest/Roost models cannot predict band gap values that are larger than the maximum band gaps in the training set \cite{xiong2020evaluating, loftis2020lattice}. Surrogate ML models with higher generalization capability are needed to future improve the prediction accuracy of our screening ML models.

Another key observation of our active learning-based screening experiments is that the machine learning models such as the graph neural network model Roost model do not perform too well on the whole dataset since the dataset is highly biased (too many data points around 0 eV). However, the ML screening model trained by the augmented samples is capable of identifying high/wide band gap materials during the screening process. The reason is that the suggested samples by the active learning tend to focus on high-bandgap materials.

We also tried to directly use active learning to search the whole design space for discovering high band gap materials, which indeed has found several new materials such as RbSr$_2$ClF$_4$ and four others. Compared to previous studies which focus on screening known materials databases, this approach has the advantage of search in the complete chemical space. However, we find that many candidates generated via this approach by the active learning model do not satisfy basic physical requirements such as neutrality and balanced electric negativity. However, these suggested samples are useful for building an ML-based screening model as they have effectively explored the design space. The resulting model can then be applied to the generated samples by MATGAN. So for semiconductor materials design, we use the MATGAN+screening approach, which is more suitable to discover semiconductors since the intuition of optimization is to find the maximum or minimum of a black-box function.

\section{Conclusion}

We propose an active learning-based generative design approach for the discovery of new inorganic materials and apply it to the discovery of high band gap materials and semiconductor materials. Our active learning-based screening is unique due to its search in the whole design space rather than known databases as done in the previous studies. Our active generative design (AGD) framework can perform an efficient guided search in the design space to identify informative samples for building screening ML models with higher accuracy compared to randomly sampled points. The ML screening model is much more computationally efficient compared to the conventional DFT-based screening model. We have successfully applied our AGD approach to the inverse design of materials with high band gap value and semiconductors and found several potential new materials, including SrYF$_5$, RbSr$_2$ClF$_4$, and seven other semiconductors as validated using DFT based principles. Our approach is thus expected to be applicable to finding materials with other properties such as thermal conductivity or superionic conductors.

\section{Contribution}
Conceptualization, J.H.; methodology, R.X. and J.H.; software, R.X.; validation, R.X., J.H.;  investigation, R.X, J.H., D.S., Y.S., Y.Z., A.N., S.L. ; resources, J.H.; writing--original draft preparation, R.X., J.H.; writing--review and editing, R.X. and J.H; visualization, R.X. and Y.S.; supervision, J.H.;  funding acquisition, J.H.

\section{Acknowledgement}
The research reported in this work was supported in part by NSF under the grant and 1940099 and 1905775 and by NSF SC EPSCoR Program under award number (NSF Award OIA-1655740 and GEAR-CRP 19-GC02). The views, perspectives, and content do not necessarily represent the official views of the National Science Foundation.

\bibliography{references}

\begin{thebibliography}{10}

\bibitem{fanourgakis2020universal}
George~S Fanourgakis, Konstantinos Gkagkas, Emmanuel Tylianakis, and George~E
  Froudakis.
\newblock A universal machine learning algorithm for large-scale screening of
  materials.
\newblock {\em Journal of the American Chemical Society}, 142(8):3814--3822,
  2020.

\bibitem{heo2019composition}
Seung~Hwae Heo, Seungki Jo, Hyo~Seok Kim, Garam Choi, Jae~Yong Song, Jun-Yun
  Kang, No-Jin Park, Hyeong~Woo Ban, Fredrick Kim, Hyewon Jeong, et~al.
\newblock Composition change-driven texturing and doping in solution-processed
  snse thermoelectric thin films.
\newblock {\em Nature communications}, 10(1):1--10, 2019.

\bibitem{oganov2019structure}
Artem~R Oganov, Chris~J Pickard, Qiang Zhu, and Richard~J Needs.
\newblock Structure prediction drives materials discovery.
\newblock {\em Nature Reviews Materials}, 4(5):331--348, 2019.

\bibitem{noh2020machine}
Juhwan Noh, Geun~Ho Gu, Sungwon Kim, and Yousung Jung.
\newblock Machine-enabled inverse design of inorganic solid materials: promises
  and challenges.
\newblock {\em Chemical Science}, 11(19):4871--4881, 2020.

\bibitem{dan2020generative}
Yabo Dan, Yong Zhao, Xiang Li, Shaobo Li, Ming Hu, and Jianjun Hu.
\newblock Generative adversarial networks (gan) based efficient sampling of
  chemical composition space for inverse design of inorganic materials.
\newblock {\em npj Computational Materials}, 6(1):1--7, 2020.

\bibitem{jain2013commentary}
Anubhav Jain, Shyue~Ping Ong, Geoffroy Hautier, Wei Chen, William~Davidson
  Richards, Stephen Dacek, Shreyas Cholia, Dan Gunter, David Skinner, Gerbrand
  Ceder, et~al.
\newblock Commentary: The materials project: A materials genome approach to
  accelerating materials innovation.
\newblock {\em Apl Materials}, 1(1):011002, 2013.

\bibitem{kirklin2015open}
Scott Kirklin, James~E Saal, Bryce Meredig, Alex Thompson, Jeff~W Doak,
  Muratahan Aykol, Stephan R{\"u}hl, and Chris Wolverton.
\newblock The open quantum materials database (oqmd): assessing the accuracy of
  dft formation energies.
\newblock {\em npj Computational Materials}, 1(1):1--15, 2015.

\bibitem{min2020accelerated}
Kyoungmin Min and Eunseog Cho.
\newblock Accelerated discovery of novel inorganic materials with desired
  properties using active learning.
\newblock {\em The Journal of Physical Chemistry C}, 2020.

\bibitem{zunger2018inverse}
Alex Zunger.
\newblock Inverse design in search of materials with target functionalities.
\newblock {\em Nature Reviews Chemistry}, 2(4):1--16, 2018.

\bibitem{tagade2019attribute}
Piyush~M Tagade, Shashishekar~P Adiga, Shanthi Pandian, Min~Sik Park,
  Krishnan~S Hariharan, and Subramanya~Mayya Kolake.
\newblock Attribute driven inverse materials design using deep learning
  bayesian framework.
\newblock {\em npj Computational Materials}, 5(1):1--14, 2019.

\bibitem{chen2020generative}
Chun-Teh Chen and Grace~X Gu.
\newblock Generative deep neural networks for inverse materials design using
  backpropagation and active learning.
\newblock {\em Advanced Science}, 7(5):1902607, 2020.

\bibitem{noh2019inverse}
Juhwan Noh, Jaehoon Kim, Helge~S Stein, Benjamin Sanchez-Lengeling, John~M
  Gregoire, Alan Aspuru-Guzik, and Yousung Jung.
\newblock Inverse design of solid-state materials via a continuous
  representation.
\newblock {\em Matter}, 1(5):1370--1384, 2019.

\bibitem{kim2020inverse}
Baekjun Kim, Sangwon Lee, and Jihan Kim.
\newblock Inverse design of porous materials using artificial neural networks.
\newblock {\em Science advances}, 6(1):eaax9324, 2020.

\bibitem{kim2018deep}
Kyungdoc Kim, Seokho Kang, Jiho Yoo, Youngchun Kwon, Youngmin Nam, Dongseon
  Lee, Inkoo Kim, Youn-Suk Choi, Yongsik Jung, Sangmo Kim, et~al.
\newblock Deep-learning-based inverse design model for intelligent discovery of
  organic molecules.
\newblock {\em npj Computational Materials}, 4(1):1--7, 2018.

\bibitem{chen2020design}
Mingkun Chen, Jiaqi Jiang, and Jonathan~A Fan.
\newblock Design space reparameterization enforces hard geometric constraints
  in inverse-designed nanophotonic devices.
\newblock {\em ACS Photonics}.

\bibitem{jiang2020multiobjective}
Jiaqi Jiang and Jonathan~A Fan.
\newblock Multiobjective and categorical global optimization of photonic
  structures based on resnet generative neural networks.
\newblock {\em Nanophotonics}, 1(ahead-of-print), 2020.

\bibitem{zhang2017computer}
Yunwei Zhang, Hui Wang, Yanchao Wang, Lijun Zhang, and Yanming Ma.
\newblock Computer-assisted inverse design of inorganic electrides.
\newblock {\em Physical Review X}, 7(1):011017, 2017.

\bibitem{elsawy2020numerical}
Mahmoud~MR Elsawy, St{\'e}phane Lanteri, R{\'e}gis Duvigneau, Jonathan~A Fan,
  and Patrice Genevet.
\newblock Numerical optimization methods for metasurfaces.
\newblock {\em Laser \& Photonics Reviews}, 14(10):1900445, 2020.

\bibitem{yuan2018accelerated}
Ruihao Yuan, Zhen Liu, Prasanna~V Balachandran, Deqing Xue, Yumei Zhou,
  Xiangdong Ding, Jun Sun, Dezhen Xue, and Turab Lookman.
\newblock Accelerated discovery of large electrostrains in batio3-based
  piezoelectrics using active learning.
\newblock {\em Advanced materials}, 30(7):1702884, 2018.

\bibitem{lookman2019active}
Turab Lookman, Prasanna~V Balachandran, Dezhen Xue, and Ruihao Yuan.
\newblock Active learning in materials science with emphasis on adaptive
  sampling using uncertainties for targeted design.
\newblock {\em npj Computational Materials}, 5(1):1--17, 2019.

\bibitem{settles2009active}
Burr Settles.
\newblock Active learning literature survey.
\newblock Technical report, University of Wisconsin-Madison Department of
  Computer Sciences, 2009.

\bibitem{schroder2020survey}
Christopher Schr{\"o}der and Andreas Niekler.
\newblock A survey of active learning for text classification using deep neural
  networks.
\newblock {\em arXiv preprint arXiv:2008.07267}, 2020.

\bibitem{lindenbaum2004selective}
Michael Lindenbaum, Shaul Markovitch, and Dmitry Rusakov.
\newblock Selective sampling for nearest neighbor classifiers.
\newblock {\em Machine learning}, 54(2):125--152, 2004.

\bibitem{melville2004diverse}
Prem Melville and Raymond~J Mooney.
\newblock Diverse ensembles for active learning.
\newblock In {\em Proceedings of the twenty-first international conference on
  Machine learning}, page~74, 2004.

\bibitem{allahyari2020coevolutionary}
Zahed Allahyari and Artem~R Oganov.
\newblock Coevolutionary search for optimal materials in the space of all
  possible compounds.
\newblock {\em npj Computational Materials}, 6(1):1--10, 2020.

\bibitem{louis2020graph}
Steph-Yves Louis, Yong Zhao, Alireza Nasiri, Xiran Wang, Yuqi Song, Fei Liu,
  and Jianjun Hu.
\newblock Graph convolutional neural networks with global attention for
  improved materials property prediction.
\newblock {\em Physical Chemistry Chemical Physics}, 22(32):18141--18148, 2020.

\bibitem{goodall2019predicting}
Rhys~EA Goodall and Alpha~A Lee.
\newblock Predicting materials properties without crystal structure: Deep
  representation learning from stoichiometry.
\newblock {\em arXiv preprint arXiv:1910.00617}, 2019.

\bibitem{tshitoyan2019unsupervised}
Vahe Tshitoyan, John Dagdelen, Leigh Weston, Alexander Dunn, Ziqin Rong, Olga
  Kononova, Kristin~A Persson, Gerbrand Ceder, and Anubhav Jain.
\newblock Unsupervised word embeddings capture latent knowledge from materials
  science literature.
\newblock {\em Nature}, 571(7763):95--98, 2019.

\bibitem{xie2018crystal}
Tian Xie and Jeffrey~C Grossman.
\newblock Crystal graph convolutional neural networks for an accurate and
  interpretable prediction of material properties.
\newblock {\em Physical review letters}, 120(14):145301, 2018.

\bibitem{snoek2012practical}
Jasper Snoek, Hugo Larochelle, and Ryan~P Adams.
\newblock Practical bayesian optimization of machine learning algorithms.
\newblock {\em Advances in neural information processing systems},
  25:2951--2959, 2012.

\bibitem{mockus1978application}
Jonas Mockus, Vytautas Tiesis, and Antanas Zilinskas.
\newblock The application of bayesian methods for seeking the extremum.
\newblock {\em Towards global optimization}, 2(117-129):2, 1978.

\bibitem{srinivas2009gaussian}
Niranjan Srinivas, Andreas Krause, Sham~M Kakade, and Matthias Seeger.
\newblock Gaussian process optimization in the bandit setting: No regret and
  experimental design.
\newblock {\em arXiv preprint arXiv:0912.3995}, 2009.

\bibitem{nogueira2014bayesian}
F~Nogueira.
\newblock Bayesian optimization: Open source constrained global optimization
  tool for python.
\newblock {\em URL https://github. com/fmfn/BayesianOptimization}, 2014.

\bibitem{hargreaves2020earth}
Cameron Hargreaves, Matthew Dyer, Michael Gaultois, Vitaliy Kurlin, and
  Matthew~J Rosseinsky.
\newblock The earth mover’s distance as a metric for the space of inorganic
  compositions.
\newblock 2020.

\bibitem{Vasp1}
G.~Kresse and J.~Hafner.
\newblock ab initio.
\newblock {\em Phys. Rev. B}, 47:558--561, Jan 1993.

\bibitem{Vasp2}
G.~Kresse and J.~Hafner.
\newblock ab initio.
\newblock {\em Phys. Rev. B}, 49:14251--14269, May 1994.

\bibitem{Vasp3}
J.~Furthmüller G.~Kresse.
\newblock Efficiency of ab initio total energy calculations for metals and
  semiconductors using a plane-wave basis set.
\newblock {\em Comput. Mater. Sci.}, 6:15--50, jul 1996.

\bibitem{Vasp4}
G.~Kresse and J.~Furthm\"uller.
\newblock Efficient iterative schemes for ab initio total-energy calculations
  using a plane-wave basis set.
\newblock {\em Phys. Rev. B}, 54:11169--11186, Oct 1996.

\bibitem{PAW1}
P.~E. Bl\"ochl.
\newblock Projector augmented-wave method.
\newblock {\em Phys. Rev. B}, 50:17953--17979, Dec 1994.

\bibitem{PAW2}
G.~Kresse and D.~Joubert.
\newblock From ultrasoft pseudopotentials to the projector augmented-wave
  method.
\newblock {\em Phys. Rev. B}, 59:1758--1775, Jan 1999.

\bibitem{GGA1}
John~P. Perdew, Kieron Burke, and Matthias Ernzerhof.
\newblock Generalized gradient approximation made simple.
\newblock {\em Phys. Rev. Lett.}, 77:3865--3868, Oct 1996.

\bibitem{GGA2}
John~P. Perdew, Kieron Burke, and Matthias Ernzerhof.
\newblock Generalized gradient approximation made simple [phys. rev. lett. 77,
  3865 (1996)].
\newblock {\em Phys. Rev. Lett.}, 78:1396--1396, Feb 1997.

\bibitem{Pymatgen}
Shyue~Ping Ong, William~Davidson Richards, Anubhav Jain, Geoffroy Hautier,
  Michael Kocher, Shreyas Cholia, Dan Gunter, Vincent~L. Chevrier, Kristin~A.
  Persson, and Gerbrand Ceder.
\newblock Python materials genomics (pymatgen): A robust, open-source python
  library for materials analysis.
\newblock {\em Computational Materials Science}, 68:314--319, 2013.

\bibitem{Hill_1952}
R~Hill.
\newblock The elastic behaviour of a crystalline aggregate.
\newblock {\em Proceedings of the Physical Society. Section A}, 65(5):349--354,
  may 1952.

\bibitem{Born_Criteria}
Félix Mouhat and François-Xavier Coudert.
\newblock Necessary and sufficient elastic stability conditions in various
  crystal systems.
\newblock {\em Physical Review B}, 90, 09 2014.

\bibitem{VASPKIT}
V.~Wang, N.~Xu, J.~Liu, Gang Tang, and W.~Geng.
\newblock Vaspkit: A user-friendly interface facilitating high-throughput
  computing and analysis using vasp code.
\newblock 2019.

\bibitem{xiong2020evaluating}
Zheng Xiong, Yuxin Cui, Zhonghao Liu, Yong Zhao, Ming Hu, and Jianjun Hu.
\newblock Evaluating explorative prediction power of machine learning
  algorithms for materials discovery using k-fold forward cross-validation.
\newblock {\em Computational Materials Science}, 171:109203, 2020.

\bibitem{loftis2020lattice}
Christian Loftis, Kunpeng Yuan, Yong Zhao, Ming Hu, and Jianjun Hu.
\newblock Lattice thermal conductivity prediction using symbolic regression and
  machine learning.
\newblock {\em The Journal of Physical Chemistry A}.

\end{thebibliography}
\bibliographystyle{unsrt}

\end{document}